\documentstyle[12pt]{article}

\begin{document}
\setcounter{page}{1} \pagestyle{plain} \vspace{1cm}
\begin{center}
\Large{\bf More on the Non-Gaussianity of Perturbations in a Non-Minimal Inflationary Model}\\
\small \vspace{1cm} {\bf R.
Shojaee$^{a,}$\footnote{r.shojaee@azaruniv.ac.ir}},\quad {\bf K.
Nozari$^{b,c,}$\footnote{knozari@umz.ac.ir }}\quad and \quad
{\bf F. Darabi$^{a,}$\footnote{f.darabi@azaruniv.ac.ir}}\\
\vspace{0.5cm}
$^{a}$Department of Physics, Azarbaijan Shahid Madani University,\\
P. O. Box 53714-161, Tabriz, Iran\\
$^{b}$Department of Physics, Faculty of Basic Sciences,\\
University of Mazandaran, P. O. Box 47416-95447, Babolsar, Iran\\
$^{c}$Research Institute for Astronomy and
Astrophysics of Maragha (RIAAM),\\
P. O. Box 55134-441, Maragha, Iran
\end{center}
\vspace{1.5cm}
\begin{abstract}
We study nonlinear cosmological perturbations and their possible
non-Gaussian character in an extended non-minimal inflation where
gravity is coupled non-minimally to both the scalar field and its
derivatives. By expansion of the action up to the third order, we
focus on the non-linearity and non-Gaussianity of perturbations in
comparison with recent observational data. By adopting an inflation
potential of the form $V(\phi)=\frac{1}{n}\lambda\phi^{n}$, we show
that for $n=4$, for instance, this extended model is consistent with
observation if $0.013<\lambda<0.095$ in appropriate units. By
restricting the equilateral amplitude of non-Gaussianity to the
observationally viable values,
the coupling parameter $\lambda$ is constraint to the values $\lambda<0.1$.
\end{abstract}
\vspace{1.5cm}
\newpage

\section{Introduction}

The idea of cosmological inflation is capable to address some
problems of the standard big bang theory, such as the horizon,
flatness and monopole problems. Also, it can provide a reliable
mechanism for generation of density perturbations responsible for
structure formation and therefore temperature anisotropies in Cosmic
Microwave Background (CMB)spectrum [1-8]. There are a wide variety
of cosmological inflation models where viability of their
predictions in comparison with observations makes them to be
acceptable or unacceptable (see for instance [9] for this purpose).
The simplest inflationary model is a single scalar field scenario in
which inflation is driven by a scalar field called the inflaton that
predicts adiabatic, Gaussian and scale-invariant fluctuations [10].
But, recently observational data have revealed some degrees of
scale-dependence in the primordial density perturbations. Also,
Planck team have obtained some constraints on the primordial
non-Gaussianity [11-13]. Therefore, it seems that extended models of
inflation which can explain or address this scale-dependence and
non-Gaussianity of perturbations are more desirable. There are a lot
of studies in this respect, some of which can be seen in Refs.
[14-19] with references therein. Among various inflationary models,
the non-minimal models have attracted much attention. Non-minimal
coupling of the inflaton field and gravitational sector is
inevitable from the renormalizability of the corresponding field
theory (see for instance [20]). Cosmological inflation driven by a
scalar field non-minimally coupled to gravity are studied, for
instance, in Refs. [21-28]. There were some issues on the unitarity
violation with non-minimal coupling (see for instance, Refs.
[29-31]) which have forced researchers to consider possible coupling
of the derivatives of the scalar field with geometry [32]. In fact,
it has been shown that a model with nonminimal coupling between the
kinetic terms of the inflaton (derivatives of the scalar field) and
the Einstein tensor preserves the unitary bound during inflation
[33]. Also, the presence of nonminimal derivative coupling is a
powerful tool to increase the friction of an inflaton rolling down
its own potential [33]. Some authors have considered the model with
this coupling term and have studied the early time accelerating
expansion of the universe as well as the late time dynamics [34-36].
In this paper we extend the non-minimal inflation models to the case
that a canonical inflaton field is coupled non-minimally to the
gravitational sector and in the same time the derivatives of the
field are also coupled to the background geometry (Einstein's
tensor). This model provides a more realistic framework for treating
cosmological inflation in essence. We study in details the
cosmological perturbations and possible non-Gaussianities in the
distribution of these perturbations in this non-minimal inflation.
We expand the action of the model up to the third order and compare
our results with observational data from Planck2015 to see the
viability of this extended model. In this manner we are able to
constraint parameter space of the model in comparison with
observation.

\section{Field Equations}

We consider an inflationary model where both a canonical scalar field and its
derivatives are coupled non-minimally to gravity. The four-dimensional action for
this model is given by the following expression:

\begin{equation}
S=\frac{1}{2}\int
d^{4}x\sqrt{-g}\Bigg[M_{p}^{2}f(\phi)R+\frac{1}{\widetilde{M}^{2}}G_{\mu\nu}\partial^{\mu}\phi\partial^{\nu}\phi-2V(\phi)\Bigg]\,,
\end{equation}
where $M_{p}$ is a reduced planck mass, $\phi$ is a canonical scalar field,
$f(\phi)$ is a general function of the scalar field and
$\widetilde{M}$ is a mass parameter. The energy-momentum tensor is obtained from action (1) as follows

\vspace{0.5cm}

$T_{\mu\nu}=\frac{1}{2\widetilde{M}^{2}}\bigg[\nabla_{\mu}\nabla_{\nu}(\nabla^{\alpha}\phi\nabla_{\alpha}\phi)
-g_{\mu\nu}\Box(\nabla^{\alpha}\phi\nabla_{\alpha}\phi)
+g_{\mu\nu}g^{\alpha\rho}g^{\beta\lambda}\nabla_{\rho}\nabla_{\lambda}(\nabla_{\alpha}\phi\nabla_{\beta}\phi)$
\begin{equation}
+\Box(\nabla_{\mu}\phi
\nabla_{\nu}\phi)\bigg]-\frac{g^{\alpha\beta}}{\widetilde{M}^{2}}
\nabla_{\beta}\nabla_{\mu}(\nabla_{\alpha}\phi
\nabla_{\nu}\phi)-M_{p}^{2}\nabla_{\mu}\nabla_{\nu}f(\phi)+M_{p}^{2}g_{\mu\nu}\Box
f(\phi)+g_{\mu\nu}V(\phi)\,.
\end{equation}

On the other hand, variation of the action (1) with respect to the
scalar field gives the scalar field equation of motion as

\begin{equation}
\frac{1}{2}M_{p}^{2}Rf'(\phi)-\frac{1}{\widetilde{M}^{2}}G^{\mu\nu}\nabla_{\mu}\nabla_{\nu}\phi-V'(\phi)=0\,,
\end{equation}
where a prime denotes derivative with respect to the scalar field. We consider a spatially flat
Friedmann-Robertson-Walker (FRW) line element as

\begin{equation}
ds^{2}=-dt^{2}+a^{2}(t)\delta_{ij}dx^{i}dx^{j}\,,
\end{equation}
where $a(t)$ is scale factor. Now, let's assume that $f(\phi)=\frac{1}{2}\phi^{2}$. In this
framework, $T_{\mu\nu}$ leads to the following energy density and
pressure for this model respectively

\begin{equation}
\rho=\frac{9H^{2}}{2\widetilde{M}^{2}}\dot{\phi}^{2}-\frac{3}{2}M_{p}^{2}H\phi(2\dot{\phi}+H\phi)+V(\phi)
\end{equation}

$$p=-\frac{3}{2}\frac{H^{2}\dot{\phi}^{2}}{\widetilde{M}^{2}}-\frac{\dot{\phi}^{2}\dot{H}}{\widetilde{M}^{2}}
-\frac{2H}{\widetilde{M}^{2}}\dot{\phi}\ddot{\phi}$$
\begin{equation}
+\frac{1}{2}M_{p}^{2}\Bigg[2\dot{H}\phi^{2}+3H^{2}\phi^{2}
+4H\phi\dot{\phi}+2\phi\ddot{\phi}+2\dot{\phi}\Bigg]-V(\phi)\,,
\end{equation}
where a dot refers to derivative with respect to the cosmic time. The equations of motion following from action (1) are

\begin{equation}
H^{2}=\frac{1}{3M_{p}^{2}}\Bigg[-\frac{3}{2}M_{p}^{2}H\phi(2\dot{\phi}+H\phi)+\frac{9H^{2}}{2\widetilde{M}^{2}}\dot{\phi}^{2}+V(\phi)\Bigg]\,,
\end{equation}

$$\dot{H}=-\frac{1}{2M_{p}^{2}}\Bigg[\dot{\phi}^{2}\bigg(\frac{3H^{2}}{\widetilde{M}^{2}}-\frac{\dot{H}}{\widetilde{M}^{2}}\bigg)
-\frac{2H}{\widetilde{M}^{2}}\dot{\phi}\ddot{\phi}-\frac{3}{2}M_{p}^{2}H\phi(2\dot{\phi}+H\phi)$$
\begin{equation}+\frac{1}{2}M_{p}^{2}\bigg((2\dot{H}+3H^{2})\phi^{2}+4H\phi\dot{\phi}+2\phi\ddot{\phi}+2\dot{\phi}^{2}\bigg)\Bigg]
\end{equation}

\begin{equation}
-3M_{p}^{2}(2H^{2}+\dot{H})\phi+\frac{3H^{2}}{\widetilde{M}^{2}}\ddot{\phi}+3H\bigg(\frac{3H^{2}}{\widetilde{M}^{2}}
+\frac{2\dot{H}}{\widetilde{M}^{2}}\bigg)\dot{\phi}+V'(\phi)=0\,.
\end{equation}

The slow-roll parameters in this model are defined as
\begin{equation}
\epsilon\equiv-\frac{\dot{H}}{H^{2}}\,\,\,\,,\,\,\,\,\eta\equiv-\frac{1}{H}\frac{\ddot{H}}{\dot{H}}\,.
\end{equation}
To have inflationary phase, $\epsilon$ and $\eta$ should satisfy slow-roll conditions($\epsilon\ll1$ , $\eta\ll1$). In our setup, we
find the following result
\begin{equation}
\epsilon=\bigg[1+\frac{\phi^{2}}{2}-\frac{\dot{\phi^{2}}}{2\widetilde{M}^{2}M_{p}^{2}}\bigg]^{-1}
\bigg[\frac{3\dot{\phi}^{2}}{2\widetilde{M}^{2}M_{p}^{2}}+\frac{\phi\dot{\phi}}{2H}
+\frac{\ddot{\phi}}{H\dot{\phi}}\bigg(\frac{\phi\dot{\phi}}{2H}
-\frac{\dot{\phi^{2}}}{\widetilde{M}^{2}M_{p}^{2}}\bigg)\bigg]
\end{equation}
and
\begin{equation}
\eta=-2\epsilon-\frac{\dot{\epsilon}}{H\epsilon}\,.
\end{equation}

Within the slow-roll approximation, equations (7),(8) and (9) can be
written respectively as
\begin{equation}
H^{2}\simeq\frac{1}{3M_{p}^{2}}\Bigg[-\frac{3}{2}M_{p}^{2}H^{2}\phi^{2}+V(\phi)\Bigg]\,,
\end{equation}

\begin{equation}
\dot{H}\simeq-\frac{1}{2M_{p}^{2}}\Bigg[\frac{3H^{2}\dot{\phi}^{2}}{\widetilde{M}^{2}}-M_{p}^{2}H\phi\dot{\phi}+M_{p}^{2}\dot{H}\phi^{2}\Bigg]\,,
\end{equation}
and
\begin{equation}
-6M_{p}^{2}H^{2}\phi+\frac{9H^{3}\dot{\phi}}{\widetilde{M}^{2}}+V'(\phi)\simeq0\,.
\end{equation}
The number of e-folds during inflation is defined as
\begin{equation}
{\cal N}=\int_{t_{hc}}^{t_{e}}H\,dt\,,
\end{equation}
where $t_{hc}$ and $t_{e}$ are time of horizon crossing and end of
inflation respectively. The number of e-folds in the slow-roll approximation in our setup can be
expressed as follows

\begin{equation}
{\cal N}\simeq\int_{\phi_{hc}}^{\phi_{e}}\frac{V(\phi)d\phi}{M_{p}^{2}\bigg(1
+\frac{1}{2}\phi^{2}\bigg)\Bigg[2M_{p}^{2}\widetilde{M}^{2}\phi
-M_{p}^{2}\widetilde{M}^{2}\frac{V'(\phi)}{V(\phi)}\bigg(1+\frac{1}{2}\phi^{2}\bigg)\Bigg]}\,.
\end{equation}
After providing the basic setup of the model, for testing cosmological viability
of this extended model we treat the perturbations in comparison with observation.

\section{Second-Order Action: Linear Perturbations}

In this section, we study linear perturbations around the
homogeneous background solution. To this end, the first step is
expanding the action (1) up to the second order in small fluctuations. It
is convenient to work in the ADM formalism given by [37]
\begin{equation}
ds^{2}=-N^{2}dt^{2}+h_{ij}(N^{i}dt+dx^{i})(N^{j}dt+dx^{j})\,,
\end{equation}
where $N^{i}$ is the shift vector and $N$ is the lapse function.
We expand the lapse function and shift vector to $N=1+2\Phi$ and
$N^{i}=\delta^{ij}\partial_{j}\Upsilon$ respectively, where $\Phi$
and $\Upsilon$ are three-scalars. Also,
$h_{ij}=a^{2}(t)[(1+2\Psi)\delta_{ij}+\gamma_{ij}]$, where $\Psi$ is
spatial curvature perturbation and $\gamma_{ij}$ is shear
three-tensor which is traceless and symmetric. In the rest of our study, we
choose $\delta\Phi=0$ and $\gamma_{ij}=0$. By taking into account
the scalar perturbations in linear-order, the metric (18) is written
as (see for instance [38])
\begin{equation}
ds^{2}=-(1+2\Phi)dt^{2}+2\partial_{i}\Upsilon
dtdx^{i}+a^{2}(t)(1+2\Psi)\delta_{ij}dx^{i}dx^{j}\,.
\end{equation}

Now by replacing metric (19) in action (1) and expanding the action up to the
second-order in perturbations, we find (see for instance [39,40])

$$S^{(2)}=\int dt dx^{3}a^{3}\Bigg[-\frac{3}{2}(M_{p}^{2}\phi^{2}-\frac{\dot{\phi}^{2}}{\widetilde{M}^{2}})\dot{\Psi}^{2}
+\frac{1}{a^{2}}((M_{p}^{2}\phi^{2}-\frac{\dot{\phi}^{2}}{\widetilde{M}^{2}})\dot{\Psi}$$
$$-(M_{p}^{2}H\phi^{2}+M_{p}^{2}\phi\dot{\phi}-\frac{3H\dot{\phi}^{2}}{\widetilde{M}^{2}})\Phi)\partial^{2}\Upsilon
-\frac{1}{a^{2}}(M_{p}^{2}\phi^{2}-\frac{\dot{\phi}^{2}}{\widetilde{M}^{2}})\Phi\partial^{2}\Psi$$
$$+3(M_{p}^{2}H\phi^{2}+M_{p}^{2}\phi\dot{\phi}-\frac{3H\dot{\phi}^{2}}{\widetilde{M}^{2}})\Phi\dot{\Psi}
+3H(-\frac{1}{2}M_{p}^{2}H\phi^{2}-M_{p}^{2}\phi\dot{\phi}$$
\begin{equation}
+\frac{3H\dot{\phi}^{2}}{\widetilde{M}^{2}})\Phi^{2}
+\frac{1}{2a^{2}}(M_{p}^{2}\phi^{2}+\frac{\dot{\phi}^{2}}{\widetilde{M}^{2}})(\partial\Psi)^{2}\Bigg]\,.
\end{equation}

By variation of action (20) with respect to $N$ and $N^{i}$ we find
\begin{equation}
\Phi=\frac{M_{p}^{2}\phi^{2}-\frac{\dot{\phi}^{2}}{\widetilde{M}^{2}}}{M_{p}^{2}H\phi^{2}+M_{p}^{2}\phi\dot{\phi}-\frac{3H\dot{\phi}^{2}}{\widetilde{M}^{2}}}\dot{\Psi}\,,
\end{equation}

$$\partial^{2}\Upsilon=\frac{2a^{2}}{3}\frac{(-\frac{9}{2}M_{p}^{2}H^{2}\phi^{2}-9M_{p}^{2}H\phi\dot{\phi}
+\frac{27H^{2}\dot{\phi}^{2}}{\widetilde{M}^{2}})}{(M_{p}^{2}H\phi^{2}+M_{p}^{2}\phi\dot{\phi}-\frac{3H\dot{\phi}^{2}}{\widetilde{M}^{2}})}$$
\begin{equation}
+3\dot{\Psi}a^{2}-\frac{M_{p}^{2}\phi^{2}-\frac{\dot{\phi}^{2}}{\widetilde{M}^{2}}}{M_{p}^{2}H\phi^{2}+M_{p}^{2}\phi\dot{\phi}-\frac{3H\dot{\phi}^{2}}{\widetilde{M}^{2}}}\dot{\Psi}\,. \end{equation}
Finally the second order action can be rewritten as follows
\begin{equation}
S^{(2)}=\int dt
dx^{3}a^{3}\vartheta_{s}\bigg[\dot{\Psi}^{2}-\frac{c_{s}^{2}}{a^{2}}(\partial\Psi)^{2}\bigg]
\end{equation}
where by definition
\begin{equation}
\vartheta_{s}\equiv6\frac{(M_{p}^{2}\phi^{2}-\frac{\dot{\phi}^{2}}{\widetilde{M}^{2}})^{2}(-\frac{1}{2}M_{p}^{2}H^{2}\phi^{2}-M_{p}^{2}H\phi\dot{\phi}+\frac{3}{\widetilde{M}^{2}}
H^{2}\dot{\phi}^{2})}{(M_{p}^{2}H\phi^{2}+M_{p}^{2}\phi\dot{\phi}-\frac{3}{\widetilde{M}^{2}}H\dot{\phi}^{2})^{2}}+
3(\frac{1}{2}M_{p}^{2}\phi^{2}-\frac{1}{2\widetilde{2}}\dot{\phi}^{2})
\end{equation}
and
$$c_{s}^{2}\equiv\frac{3}{2}\bigg\{(M_{p}^{2}\phi^{2}-\frac{\dot{\phi}^{2}}{\widetilde{M}^{2}})^{2}
(M_{p}^{2}H\phi^{2}+M_{p}^{2}\phi\dot{\phi}-\frac{3H\dot{\phi}^{2}}{\widetilde{M}^{2}})H$$
$$-(M_{p}^{2}H\phi^{2}+M_{p}^{2}\phi\dot{\phi}-\frac{3H\dot{\phi}^{2}}{\widetilde{M}^{2}})^{2}(M_{p}^{2}\phi^{2}-\frac{\dot{\phi}^{2}}{\widetilde{M}^{2}})$$
$$4(M_{p}^{2}\phi^{2}-\frac{\dot{\phi}^{2}}{\widetilde{M}^{2}})(M_{p}^{2}\phi\dot{\phi}-\frac{\dot{\phi}\ddot{\phi}}{\widetilde{M}^{2}})
(M_{p}^{2}H\phi^{2}+M_{p}^{2}\phi\dot{\phi}-\frac{3H\dot{\phi}^{2}}{\widetilde{M}^{2}})$$
$$-(M_{p}^{2}-\frac{\dot{\phi}^{2}}{\widetilde{M}^{2}})^{2}(M_{p}^{2}\dot{H}\phi^{2}+2M_{p}^{2}H\phi\dot{\phi}M_{p}^{2}\dot{\phi}^{2}+M_{p}^{2}\phi\ddot{\phi}
-\frac{3\dot{H}\dot{\phi}^{2}}{\widetilde{M}^{2}}-\frac{6}{\widetilde{M}^{2}}H\dot{\phi}\ddot{\phi})\bigg\}$$
$$\bigg\{9[\frac{1}{2}M_{p}^{2}\phi^{2}-\frac{\dot{\phi}^{2}}{2\widetilde{M}^{2}}][4(\frac{1}{2}M_{p}^{2}\phi^{2}-\frac{\dot{\phi}^{2}}{2\widetilde{M}^{2}})
(-\frac{1}{2}M_{p}^{2}H^{2}\phi^{2}-M_{p}^{2}H\phi\dot{\phi}+\frac{3}{M\widetilde{^{2}}H^{2}\dot{\phi}^{2}})$$
\begin{equation}
+(M_{p}^{2}H\phi^{2}+M_{p}^{2}\phi\dot{\phi}-\frac{3H\dot{\phi}^{2}}{\widetilde{M}^{2}})^{2}]\bigg\}^{-1}\,.
\end{equation}

In order to obtain quantum perturbations $\Psi$, we can find
equation of motion of the curvature perturbation by varying action (23)
which follows
\begin{equation}
\ddot{\Psi}+\bigg(3H+\frac{\dot{\vartheta_{s}}}{\vartheta_{s}}\bigg)+\frac{c_{s}^{2}k^{2}}{a^{2}}\Psi=0\,.
\end{equation}
By solving the above equation up to the lowest order in slow-roll approximation,
we find
\begin{equation}
\Psi=\frac{iH\exp(-ic_{s}k\tau)}{2c_{s}^{\frac{3}{2}}\sqrt{k^{3}}\vartheta_{s}}(1+ic_{s}k\tau)\,.
\end{equation}
By using the two-point correlation functions we can study power
spectrum of curvature perturbation in this setup. We find two-point correlation
function by obtaining vacuum expectation value at the end of inflation.
We define the power spectrum $P_{s}$, as
\begin{equation}
\langle0|\Psi(0,\textbf{k}_{1})\Psi(0,\textbf{k}_{2})|0\rangle=\frac{2\pi^{2}}{k^{3}}P_{s}(2\pi)^{3}\delta^{3}(\textbf{k}_{1}+\textbf{k}_{2})\,,
\end{equation}
where
\begin{equation}
P_{s}=\frac{H^{2}}{8\pi^{2}\vartheta_{s} c_{s}^{3}}\,.
\end{equation}

The spectral index of scalar perturbations is given by (see Refs. [41-43] for more details on the cosmological perturbations in generalized gravity theories and also inflationary spectral index in these theories.)

\begin{equation}
n_{s}-1=\frac{d\ln P_{s}}{d\ln
k}|_{c_{s}k=aH}=-2\epsilon-\delta_{F}-\eta_{s}-S
\end{equation}
where by definition
\begin{equation}
\delta_{F}=\frac{\dot{f}}{H(1+f)}\,\,\,\,,\,\,\,\,\eta_{s}=\frac{\dot{\epsilon_{s}}}{H\epsilon_{s}}\,\,\,\,,\,\,\,\,S=\frac{\dot{c_{s}}}{Hc_{s}}
\end{equation}
also
\begin{equation}
\epsilon_{s}=\frac{\vartheta_{s}c_{s}^{2}}{M_{pl}^{2}(1+f)}.
\end{equation}

we obtain finally

\begin{equation}
n_{s}-1=-2\epsilon-\frac{1}{H}\frac{d\ln c_{s}}{dt}
-\frac{1}{H}\frac{d\ln[2H(1+\frac{\phi^{2}}{2})\epsilon+\phi\dot{\phi}]}{dt}\,,
\end{equation}
which shows the scale dependence of perturbations due to deviation of $n_{s}$
from $1$.

Now we study tensor perturbations in this setup. To this end, we write the metric as follows
\begin{equation}
ds^{2}=-dt^{2}+a(t)^{2}(\delta_{ij}+T_{ij})dx^{i}dx^{j}\,,
\end{equation}
where $T_{ij}$ is a spatial shear 3-tensor which is transverse and
traceless. It is convenient to write $T_{ij}$ in terms of two
polarization modes, as follows
\begin{equation}
T_{ij}=T_{+}e^{+}_{ij}+T^{\times}e^{\times}_{ij}\,,
\end{equation}
where $e^{+}_{ij}$ and $e^{\times}_{ij}$ are the polarization tensors. In this case the second order action for the tensor mode can de
written as
\begin{equation}
S_{T}=\int dt dx^{3}
a^{3}\vartheta_{T}\bigg[\dot{T}_{(+,\times)}^{2}-\frac{c_{T}^{2}}{a^{2}}(\partial
T_{(+,\times)})^{2}\bigg]\,,
\end{equation}
where by definition
\begin{equation}
\vartheta_{T}\equiv\frac{1}{8}(M_{p}^{2}\phi^{2}-\frac{\dot{\phi}^{2}}{\widetilde{M}^{2}})
\end{equation}
and
\begin{equation}
c_{T}^{2}\equiv\frac{\widetilde{M}^{2}M_{p}^{2}\phi^{2}+\dot{\phi}^{2}}{\widetilde{M}^{2}M_{p}^{2}\phi^{2}-\dot{\phi}^{2}}\,.
\end{equation}

Now, the amplitude of tensor perturbations is given by
\begin{equation}
P_{T}=\frac{H^{2}}{2\pi^{2}\vartheta_{T}c_{T}^{3}}\,,
\end{equation}
where we have defined the tensor spectral index as
\begin{equation}
n_{T}\equiv\frac{d\ln P_{T}}{d\ln
k}|_{c_{T}k=aH}\,=-2\epsilon-\delta_{F}.
\end{equation}
By using above equations we get finally
\begin{equation}
n_{T}=-2\epsilon-\frac{\phi\dot{\phi}}{H(1+\frac{\phi^{2}}{2})}\,.
\end{equation}

The tensor-to-scalar ratio as an important observational quantity in our setup is given by
\begin{equation}
r=\frac{P_{T}}{P_{s}}=16c_{s}\bigg(\epsilon+\frac{\phi\dot{\phi}}{2H(1+\frac{\phi^{2}}{2})}+O(\epsilon^{2})\bigg)\simeq-8c_{s}n_{T}
\end{equation}
which yields the standard consistency relation.

\section{Third-Order Action: Non-Gaussianity}

Since a two-point correlation function of the scalar perturbations
gives no information about possible non-Gaussian feature of distribution, we study
higher-order correlation functions. A three-point correlation function is capable to give the required information. For this
purpose, we should expand action (1) up to the third order in small
fluctuations around the homogeneous background solutions. In this respect we obtain

\vspace{0.5cm} $S^{(3)}=\int
dtdx^{3}a^{3}\bigg\{3\Phi^{3}[M_{p}^{2}H^{2}(1+\frac{\phi^{2}}{2})+M_{p}^{2}H\phi\dot{\phi}-\frac{5}{\widetilde{M^{2}}}H^{2}\dot{\phi^{2}}]
+\Phi^{2}[9\Psi(-\frac{1}{2}M_{p}^{2}\phi^{2}-M_{p}^{2}H\phi\dot{\phi}$
\vspace{0.5cm} $+\frac{3}{\widetilde{M}^{}}H^{2}\dot{\phi}^{2})
+6\dot{\Psi}(-M_{p}^{2}H(1+\frac{\phi^{2}}{2})-\frac{1}{2}M_{p}^{2}\phi\dot{\phi}\frac{3}{\widetilde{M}^{2}}H\dot{\phi}^{2})
-\frac{\dot{\phi}^{2}}{\widetilde{M}^{2}a^{2}}\partial^{2}\Psi
-\frac{2}{a^{2}}\partial^{2}\Upsilon(-M_{p}^{2}H$\vspace{0.05cm}
$(1+\frac{\phi^{2}}{2})-\frac{1}{2}M_{p}^{2}\phi\dot{\phi}\frac{3}{\widetilde{M}^{2}}H\dot{\phi}^{2})]
+\Phi[\frac{1}{a^{2}}(-M_{p}^{2}H\phi^{2}-M_{p}^{2}\phi\dot{\phi}+\frac{3H\dot{\phi}^{2}}{\widetilde{M}^{2}})\partial_{i}\Psi\partial_{i}\Upsilon
-9(-M_{p}^{2}H\phi^{2}-M_{p}^{2}\phi\dot{\phi}+$ \vspace{0.5cm}
$\frac{3H\dot{\phi}^{2}}{\widetilde{M}^{2}})\dot{\Psi}\Psi+\frac{1}{2a^{4}}(M_{p}^{2}(1+\frac{\phi^{2}}{2})+\frac{3}{2}\frac{\dot{\phi}^{2}}{\widetilde{M}^{2}})
(\partial_{i}\partial_{j}\Upsilon\partial_{i}\partial_{j}\Upsilon-\partial^{2}\Upsilon\partial^{2}\Upsilon)
+\frac{1}{a^{2}}(-M_{p}^{2}H\phi^{2}-M_{p}^{2}\phi\dot{\phi}+\frac{3H\dot{\phi}^{2}}{\widetilde{M}^{2}})\Psi\partial^{2}\Upsilon
$ \vspace{0.5cm}
$+\frac{4}{2a^{2}}(M_{p}^{2}(1+\frac{\phi^{2}}{2})+\frac{3}{2}\frac{\dot{\phi}^{2}}{\widetilde{M}^{2}})
\dot{\Psi}\partial^{2}\Upsilon+\frac{1}{a^{2}}(-M_{p}^{2}\phi^{2}+\frac{\dot{\phi}}{\widetilde{M}^{2}})
\Psi\partial^{2}\Psi
+\frac{1}{2a^{2}}(-M_{p}^{2}\phi^{2}+\frac{\dot{\phi}}{\widetilde{M}^{2}})(\partial\Psi)^{2}
-6(M_{p}^{2}(1+\frac{\phi^{2}}{2})+$ \vspace{0.5cm}
$\frac{3}{2}\frac{\dot{\phi}^{2}}{\widetilde{M}^{2}})\dot{\Psi}^{2}]+\frac{1}{2a^{2}}(M_{p}^{2}\phi^{2}+\frac{\dot{\phi}^{2}}{\widetilde{M}^{2}})
\Psi(\partial\Psi)^{2}
+\frac{9}{2}(-M_{p}^{2}\phi^{2}+\frac{\dot{\phi}}{\widetilde{M}^{2}})\dot{\Psi^{2}}\Psi
-\frac{1}{a^{2}}(-M_{p}^{2}\phi^{2}+\frac{\dot{\phi}}{\widetilde{M}^{2}})\dot{\Psi}\partial_{i}\Psi\partial_{i}\Upsilon
-\frac{1}{a^{2}}(-M_{p}^{2}\phi^{2}+\frac{\dot{\phi}}{\widetilde{M}^{2}})$
\begin{equation}
\dot{\Psi}\Psi\partial^{2}\Upsilon-\frac{3}{4a^{4}}\Psi(-M_{p}^{2}\phi^{2}+\frac{\dot{\phi}}{\widetilde{M}^{2}})
(\partial_{i}\partial_{j}\Upsilon\partial_{i}\partial_{j}\Upsilon-\partial^{2}\Upsilon\partial^{2}\Upsilon)
+\frac{1}{a^{4}}(-M_{p}^{2}\phi^{2}+\frac{\dot{\phi}}{\widetilde{M}^{2}})\partial_{i}\Psi\partial_{i}\Upsilon\partial^{2}\Upsilon\bigg\}
\end{equation}

We use Eqs. (21) and (22) for eliminating $\Phi$ and $\Upsilon$ in this relation. For this end, we introduce the quantity $\chi$ as follows

\begin{equation}
\Upsilon=\frac{M_{p}^{2}\widetilde{M}^{2}\phi^{2}-\dot{\phi}^{2}}{\widetilde{M}^{2}M_{p}^{2}(H\phi^{2}+\phi\dot{\phi})-3H\dot{\phi}^{2}}\Psi
+\frac{2\widetilde{M^{2}}a^{2}\chi}{M_{p}^{2}\widetilde{M}^{2}\phi^{2}-\dot{\phi}^{2}}\,,
\end{equation}
where

\begin{equation}
\partial^{2}\chi=\vartheta_{s}\dot{\Psi}\,.
\end{equation}

Now the third order action (43) takes the following form

\vspace{0.5cm}

$S^{(3)}=\int dt\,
dx^{3}a^{3}\bigg\{[-3M_{p}^{2}c_{s}^{-2}\Psi\dot{\Psi^{2}}
+M_{p}^{2}a^{-2}\Psi(\partial\Psi)^{2}+M_{p}^{2}c_{s}^{-2}H^{-1}\dot{\Psi}^{3}]$
\begin{equation}
\bigg[(1+\frac{1}{4}\phi^{2})\epsilon+\frac{5}{8}\frac{\phi\dot{\phi}}{H}\bigg]
-2(1+\frac{1}{4}\phi^{2})^{-1}(\frac{5}{8}\frac{\phi\dot{\phi}}{c_{s}^{2}H})\dot{\Psi}\partial_{i}\Psi\partial_{i}\chi\bigg\}\,.
\end{equation}

By calculating the three-point correlation function we can study
non-Gaussianity feature of the primordial perturbations. For the
present model, we use the interaction picture in which the
interaction Hamiltonian, $H_{int}$, is equal to the Lagrangian third
order action. The vacuum expectation value of curvature
perturbations at $\tau=\tau_{f}$ is

\begin{equation}
\langle\Psi(\textbf{k}_{1})\Psi(\textbf{k}_{2})\Psi(\textbf{k}_{3})\rangle=-i\int_{\tau_{i}}^{\tau_{f}}d\tau
\langle0|[\Psi(\tau_{f},\textbf{k}_{1})\Psi(\tau_{f},\textbf{k}_{2})\Psi(\tau_{f},\textbf{k}_{3}),H_{int}(\tau)]|0\rangle\,.
\end{equation}

By solving the above integral in Fourier space, we find
\begin{equation}
\langle\Psi(\textbf{k}_{1})\Psi(\textbf{k}_{2})\Psi(\textbf{k}_{3})\rangle=(2\pi)^{3}\delta^{3}(\textbf{k}_{1}+\textbf{k}_{2}+\textbf{k}_{3})
P_{s}^{2}F_{\Psi}(\textbf{k}_{1},\textbf{k}_{2},\textbf{k}_{3})\,,
\end{equation}
where
\begin{equation}
F_{\Psi}(\textbf{k}_{1},\textbf{k}_{2},\textbf{k}_{3})=\frac{(2\pi)^{2}}{\prod_{i=1}^{3}k_{i}^{3}}G_{\Psi}\,,
\end{equation}

\vspace{0.5cm}

$G_{\Psi}=\bigg[\frac{3}{4}\bigg(\frac{2}{K}\Sigma_{i>j}k_{i}^{2}k_{j}^{2}-\frac{1}{K^{2}}\Sigma_{i\neq
j}k_{i}^{2}k_{j}^{3}\bigg)+\frac{1}{4}\bigg(\frac{1}{2}\Sigma_{i}k_{i}^{3}+\frac{2}{K}\Sigma_{i>j}k_{i}^{2}k_{j}^{2}
-\frac{1}{K^{2}}\Sigma_{i\neq j} k_{i}^{2}k_{j}^{3}\bigg)$
\begin{equation}
-\frac{3}{2}\bigg(\frac{(k_{1}k_{2}k_{3})^{2}}{K^{3}}\bigg)\bigg]\bigg(1-\frac{1}{c_{s}^{2}}\bigg)\,,
\end{equation}

and $K=\sum_{i}k_{i}$. Finally the non-linear parameter $f_{NL}$ is defined as follows

\begin{equation}
f_{NL}=\frac{10}{3}\frac{G_{\Psi}}{\sum_{i=1}^{3}k_{i}}\,.
\end{equation}

Here we study non-Gaussianity in the orthogonal and the equilateral
configurations [44,45]. Firstly we should account $G_{\Psi}$ in
these configurations. To this end, we follow Refs. [46-48] to
introduce a shape $\zeta_{\ast}^{equi}$ as
$\zeta_{\ast}^{equi}=-\frac{12}{13}(3\zeta_{1}-\zeta_{2})$. In this
manner we define the following shape which is orthogonal to
$\zeta_{\ast}^{equi}$
\begin{equation}
\zeta_{\ast}^{ortho}=-\frac{12}{14-13\beta}[\beta(3\zeta_{1}-\zeta_{2})+3\zeta_{1}-\zeta_{2}]\,,
\end{equation}
where $\beta\simeq1.1967996$. Finally, bispectrum (48) can be
written in terms of $\zeta_{\ast}^{equi}$ and $\zeta_{\ast}^{ortho}$
as follows
\begin{equation}
G_{\Psi}=G_{1}\zeta_{\ast}^{equi}+G_{2}\zeta_{\ast}^{ortho}\,,
\end{equation}
where
\begin{equation}
G_{1}=\frac{13}{12}\bigg[\frac{1}{24}\bigg(1-\frac{1}{c_{s}^{2}}\bigg)\bigg](2+3\beta)
\end{equation}
and
\begin{equation}
G_{2}=\frac{14-13\beta}{12}\bigg[\frac{1}{8}\bigg(1-\frac{1}{c_{s}^{2}}\bigg)\bigg]\,.
\end{equation}

Now, by using equations (50-55) we obtain the amplitude of
non-Gaussianity in the orthogonal and equilateral configurations
respectively as
\begin{equation}
f_{NL}^{equi}=\frac{130}{36\sum_{i=1}^{3}k_{i}^{3}}\bigg[\frac{1}{24}
\bigg(\frac{1}{1-c_{s}^{2}}\bigg)\bigg](2+3\beta)\zeta_{\ast}^{equi}\,,
\end{equation}
and
\begin{equation}
f_{NL}^{ortho}=\frac{140-130\beta}{36\sum_{i=1}^{3}k_{i}^{3}}\bigg[\frac{1}{8}
\bigg(1-\frac{1}{c_{s}^{2}}\bigg)\bigg]\zeta_{\ast}^{ortho}\,.
\end{equation}

The equilateral and the orthogonal shape have a negative and a positive peak in $k_{1}=k_{2}=k_{3}$ limit, respectively [49].
Thus, we can rewrite the above equations in this limit as
\begin{equation}
f_{NL}^{equi}=\frac{325}{18}\bigg[\frac{1}{24}\bigg(\frac{1}{c_{s}^{2}}-1\bigg)\bigg](2+3\beta)\,,
\end{equation}
and
\begin{equation}
f_{NL}^{ortho}=\frac{10}{9}\bigg[\frac{1}{8}\bigg(1-\frac{1}{c_{s}^{2}}\bigg)\bigg](\frac{7}{6}+\frac{65}{4}\beta)\,,
\end{equation}
respectively.

\section{Confronting with Observation}

The previous sections were devoted to the theoretical framework of
this extended model. In this section we compare our model with
observational data to find some observational constraints on the
model parameter space. In this regard, we introduce a suitable
candidate for potential term in the action. We adopt\footnote{Note that in general
$\lambda$ has dimension related to the Planck mass. This can be seen easily by considering the normalization of $\phi$ via
$V(\phi)=\frac{1}{n}\lambda(\frac{\phi}{\phi_{0}})^{n}$ which indicates that $\lambda$ cannot be dimensionless in general. When we consider some numerical values for $\lambda$ in our numerical analysis, these values are in \emph{``appropriate units"}.}
$V(\phi)=\frac{1}{n}\lambda\phi^{n}$ which contains some interesting
inflation models such as chaotic inflation. To be more specified, we
consider a quartic potential with $n=4$. Firstly we substitute this
potential into equation (11) and then by adopting $\epsilon=1$ we
find the inflaton field's value at the end of inflation. Then by
solving the integral (17), we find the inflaton field's value at the
horizon crossing in terms of number of e-folds, $N$. Then we
substitute $\phi_{hc}$ into Eqs. (33), (42), (58) and (59). The
resulting relations are the basis of our numerical analysis on the
parameter space of the model at hand. To proceed with numerical
analysis, we study the behavior of the tensor-to-scalar ratio versus
the scalar spectral index. In figure (1), we have plotted the
tensor-to-scalar ratio versus the scalar spectral index for $N=60$
in the background of Planck2015 data. The trajectory of result in
this extended non-minimal inflationary model lies well in the confidence levels
of Planck2015 observational data for viable spectral index and $r$.
The amplitude of orthogonal configuration of non-Gaussianity versus
the amplitude of equilateral configuration is depicted in figure 2
for $N=60$. We see that this extended non-minimal model, in some
ranges of the parameter $\lambda$, is consistent with observation.
If we restrict the spectral index to the observationally viable
interval $0.95<n_{s}<0.97$, then $\lambda$ is constraint to be in
the interval $0.013<\lambda<0.095$ in \emph{appropriate units}. If we
restrict the equilateral configuration of non-Gaussianity to the
observationally viable condition $-147<f^{equi}_{NL}<143$, then we
find the constraint $\lambda<0.1$ in our setup.

\vspace{1cm}

\begin{figure}[htp]
\begin{center}\includegraphics{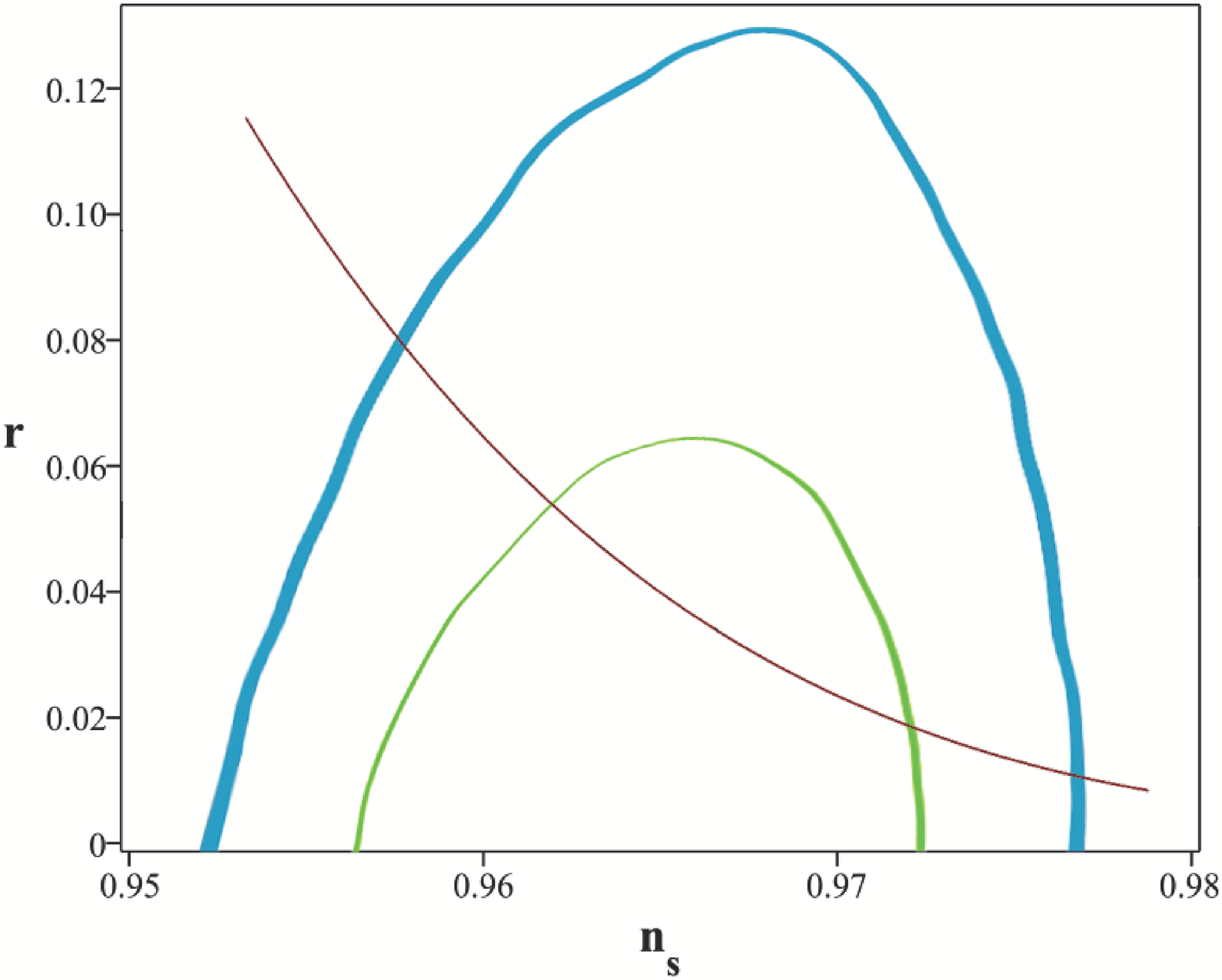} \vspace{6cm}
\end{center}
\caption{\small {Tensor-to-scalar ratio versus the scalar spectral
index in the background of Planck2015 TT,TE and EE+lowP data.}}
\end{figure}

\begin{figure}[htp]
\begin{center}\includegraphics{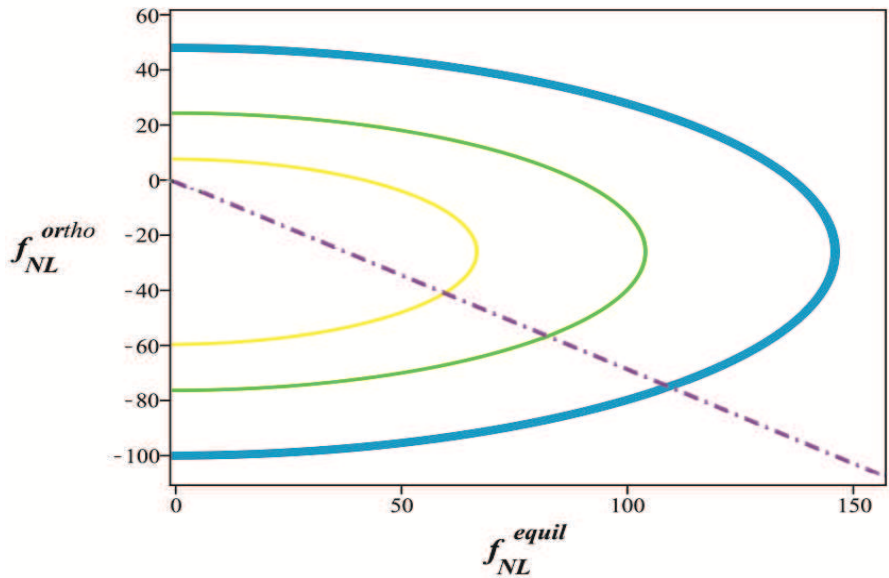} \vspace{5.75cm}
\end{center}
\caption{\small {The amplitude of the orthogonal configuration
versus the amplitude of the equilateral configuration of non-Gaussianity in the background of Planck2015 TTT, EEE, TTE and
EET data.}}
\end{figure}

\section{Summary and Conclusion}

We studied an extended model of single field inflation where the inflaton
and its derivatives are coupled to the background geometry.
By focusing on the third order action and nonlinear perturbations we
obtained observables of cosmological inflation, such as tensor-to-scalar
ratio and the amplitudes of non-Gaussianities in this extended setup. By confronting
the model's outcomes with observational data from Planck2015, we were able to constraint
parameter space of the model. By adopting a quartic potential with
$V(\phi)=\frac{1}{4}\lambda\phi^{4}$, restricting the model to realize
observationally viable spectral index (or tensor-to-scalar ratio)
imposes the constraint on coupling $\lambda$ as $0.013<\lambda<0.095$.
Also restricting the amplitude of equilateral amplitude of non-Gaussianity
to the observationally supported value of $-147<f^{equi}_{NL}<143$, results in the constraint
$\lambda<0.1$ in appropriate units. \\

{\bf Acknowledgement}\\
The work of K. Nozari has been supported financially by Research
Institute for Astronomy and Astrophysics of Maragha (RIAAM) under
research project number 1/5750-1.

\end{document}